# TIME-VARYING LINEAR MODEL APPROXIMATION: APPLICATION TO THERMAL AND AIRFLOW BUILDING SIMULATION

Thierry BERTHOMIEU, Harry BOYER
Laboratoire de Génie Industriel – Université de la Réunion

## ABSTRACT

Considering the natural ventilation, the thermal behavior of buildings can be described by a linear time varying model. In this paper, we describe an implementation of model reduction of linear time varying systems.

We show the consequences of the model reduction on computing time and accuracy.

Finally, we compare experimental measures and simulation results using the initial model or the reduced model. The reduced model shows negligible difference in accuracy, and the computing time shortens.

## INTRODUCTION

Nowadays, most of the numerical tools dedicated to simulating the thermal behavior of buildings, consider a lot of physical phenomena : airflow, humidity transfers… They can solve miscellaneous problems for very different users : designers, architecs or researchers. They handle complex multi-zones buildings and automatically generate the very detailed numeric models.

The complexity of the problems grows simultaneously with the calculation possibilities. Actually, the power of the available computers is a limitation. Thus, to reduce the computing time is still an open challenge.

After spacial discretisation, the thermal model of a building is a large linear system. Robust and accurate methods of model reduction are available for time-invariant systems (Menezo,1998)(Déqué, 1997).

But, in the particular case of airflow taken into account, the thermal model becomes a time varying system because of the varying airflow rates between zones.

CODYRUN is a software dedicated to thermal simulation, including natural ventilation and humidity transfers (Boyer, 1996, 1998, 1999)(Garde, 2000). The paper is centered on the integration of balanced reduction routines within CODYRUN, in the particular context of a time varying model.

The effectiveness of the approach is demonstrated by its application to the simulation of a multi-zones building.

## THERMAL AND AIRFLOW MODELS

A building may be decomposed into several zones. The thermal behavior of each zone is homogeneous, and described by a differential system :

- Equations with constant coefficients describe the evolution of the field of temperatures in walls and glazings.

- One equation describes the evolution of the dry air temperature. Its coefficients are not constant if airflow rates are time dependant.

Written in matrix form, the following linear equation describes the thermal behaviour of a zone :

$$\dot{T} = A T + B u \qquad (1)$$

Where   T is the vector of the nodal temperatures u is the vector of the applied sollicitations.

This system is time-invariant if the convective coefficients are not time dependant and if airflow is neglected or constant.

Thereafter, we will take account of the airflow, and $A(t)$ and $B(t)$ will be time-varying matrices.

$$\dot{T} = A(t) T + B(t)u \qquad (2)$$

In CODYRUN, $B(t)$ and u was framed together as a single vector $B(t)u$.

In order to implement the classical numerical schemes of reduction, our first step was to dissociate $B(t)$ and u in the standard form (2).

We chose for u the following frame :

- meteorological sollicitations

- short wave flux densities on envelopes surfaces

- air contribution

- variables for thermal coupling between zones

The order the system (2) is usually about a few dozen. Users often need to calculate the variables (temperatures, heat power) every hour, over a season or a year : model reduction may efficiently reduce the computing time.

Indeed, at every step of time, CODYRUN solves apart two models (thermal and airflow). The two modules are coupled, thanks to an iterative algorithm. The coupling variables are the air mass flows.

The airflow model is non linear and based on pressure variables. It take into account the principal driving effects : the wind and the thermal buoyancy. It allows the determination of the airflow network in the building (Boyer, 1999). Reduction of the thermal model has no incidence on the equations of the airflow model.

Our approach for reducing the computing time is to reduce the order of the thermal model. Balanced model reduction was originally developed for time-invariant models. In the next section, we briefly describe the balanced model reduction process.

## BALANCED REDUCTION OF THERMAL MODEL

Equation (2) is transformed by a change of coordinates to the new state space formulation (Gille, 1984) :

$\dot{X} = [M^{-1}AM] X + [M^{-1}B] u$

$T = [M] X$

Where M is the coordinate transformation matrix,

$X(t)$ is the new state vector,

$u(t)$ the input and $T(t)$ the output of the system.

The problem is to extract a subsystem of this full order system which works on that part of the state space which is most involved in the input-output behavior of the original system. Two matrices (called the observability Gramian and the controllability Gramian) enable to identify which part of the state space is controlled most strongly and which part is most strongly observed.

It is possible to construct a coordinate transformation matrix M so that the two gramians become equal and diagonal. The diagonal terms are the Hankel singular values which evaluate the observability and the controllability of each state variable. The resulting state model is a "balanced realization".

The reduced order model is obtained by extracting a subsystem from the balanced realization (Moore, 1981)(Tombs, 1987) :

$\dot{X}_r = A_r X_r + B_r u$

$T = C_r X_r + D_r u$

$X_r(t)$ is the reduced state vector.

Notice that the matrix $D_r$ must preserve the static gain of the original system.

We have implemented in CODYRUN balanced reduction tool that is available in the numerical library SLICOT (Varga, 2002).

We describe in the next sections several methods for adapting this numerical tool to our time-varying case.

## MODEL REDUCTION OF TIME-VARYING SYSTEMS

To compute the reduced order model requires more operations than to solve the original system. Thus, considering a time-varying system, model reduction cannot be achieved at each step of time.

We found two solutions for this problem :

**conditional model reduction**

When the global model of the building is not very sensitive to airflow (closed building, known flow rates …), the reduced order model remains a good approximation during a large time of simulation.

We need a criterion of precision in order to test the validity of the initial model. This criterion determines the updating of the reduced order model when necessary.

For exemple, we compute the reduced order model, when variation of one inter-zone airflow exceeds a given tolerance.

For implementation, notice that the numerical differentiation algorithm uses $X_r(k-1)$ – the state vector at previous step of time - to compute the current vector $X_r(k)$.

Every time the model is updated, $X_r(k-1)$ is no longer valid and must be estimated from sollicitations $u(k-1)$ and temperatures $T(k-1)$ at previous step.

We have $T^{(k-1)} = C_r X_r^{(k-1)} + D_r u^{(k-1)}$

Thus, minimize $\| C_r X_r^{(k-1)} - (T^{(k-1)} - D_r u^{(k-1)}) \|_2$ give $X_r(k-1)$. This is a classical least squares problem.

**separate model reduction**

In our case, most of the equations of the thermal system have constant coefficients. Thus, we previously separate the time-invariant part. This method yields a time-invariant sub-system which can be reduced only one time.

In the differential system (2), let $X_1$ be the vector of temperatures of the envelopes and $X_2$ the dry temperature of the zone.

$$\begin{bmatrix} \dot{X}_1 \\ \dot{X}_2 \end{bmatrix} = \begin{bmatrix} A_{11} & A_{12} \\ A_{21}(t) & A_{22}(t) \end{bmatrix} \begin{bmatrix} X_1 \\ X_2 \end{bmatrix} + \begin{bmatrix} B_1 \\ B_2(t) \end{bmatrix} u$$

$A_{11}$, $A_{12}$, et $B_1$ are constants,

$A_{21}$, $A_{22}$, et $B_2$ are time-varying terms.

To obtain the equation of evolution of $X_1$ we extend the input vector u with an estimation $\hat{X}_2$ of $X_2$

$$\dot{X}_1 = A_{11} X_1 + [B_1 \mid A_{12}] \begin{pmatrix} u \\ \hat{X}_2 \end{pmatrix} \quad (3)$$

Model reduction of this time-invariant linear system presents no difficulty and is computed only once.

To obtain the evolution equation of $X_2$ we extend the input vector u with an estimation $\hat{X}_1$ of $X_1$

$$\dot{X}_2 = A_{22}(t).X_2 + [B_2(t) \mid A_{21}(t)] \begin{pmatrix} u \\ \hat{X}_1 \end{pmatrix} \quad (4)$$

It is necessary to refine the estimation $\hat{X}_2$, in an iterative way : the result X2 of the equation (4) is returned in input of the equation (3), until $\|\hat{X}_2 - X_2\| < \varepsilon$ given.

Figure 1 represents the partitioned system

Figure 2 represents the algorithm using the reduced order model.

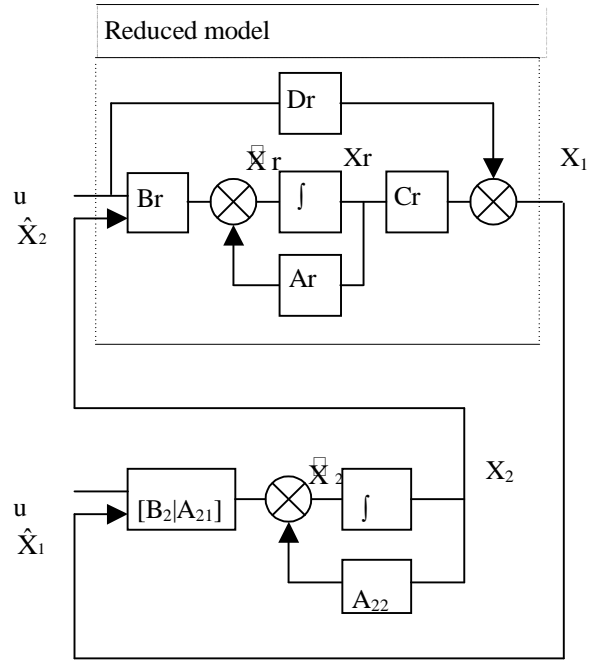

Figure 2 : *reduced model*

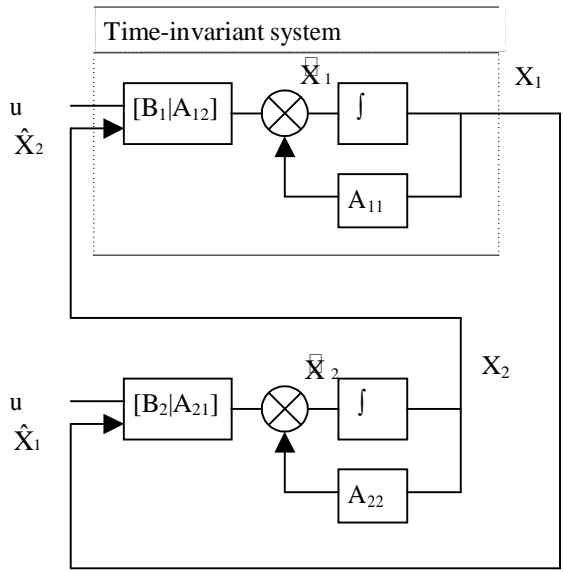

Figure 1 : *partition of state variables*

## REDUCTION ORDER

A building is composed of thermal zones, which are described by systems of different orders ni. We have to choose a set of reduced order nri. The computing time and the precision of the simulation depend on this choice.

Notice that the computing time of a simulation is about :

$$t_g = t_f + c. \sum_{every\ zones} nr_i^3$$

where tf and c are constants.

Balanced reduction have explicit error bounds :

Let [G] and [Gr] be the transfer-function matrices of the original and reduced systems respectively.

We have $\| [G] - [Gr] \|_\infty \leq 2 * (\sigma_{nr+1} + \ldots + \sigma_n)$

where $\sigma_1 \geq \sigma_2 \ldots \geq \sigma_n \geq 0$ are the Hankel singular values.

Thus $\sigma_{nr} > \varepsilon$ can be a useful criterion to automatically select a reduced order nr.

If the same precision is required for every zones, the same tolerance $\varepsilon$ can be applied to every system.

Figure 3 shows the performances of the reduced order model of a 5-zones closed building.

For some values of the tolerance ε, the figure shows :

- the computing time of a simulation (relatively to the computing time with the initial system)
- the reduction error, ie the difference between the temperatures computed with the initial model and the reduced model (standard deviation)

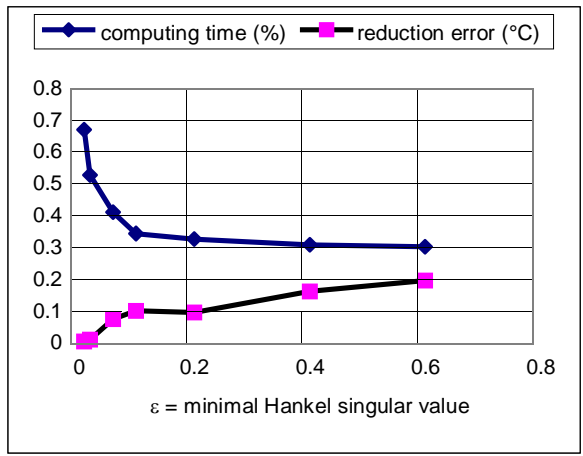

Fig. 3 : *Performance of reduced model*

A compromise time/precision appears and we find no necessity of a severe reduction.

## EXPERIMENTATION

Our goal is not to validate the software CODYRUN. This validation is the purpose of many previous works (Garde, 2001).

We will compare measured and simulated temperatures in order to relativize the differences between the initial model and the reduced model.

Figure 4 shows a typical dwelling of collective building in Reunion Island. It includes three bedrooms and a living room.

A measurement campaign took place in this dwelling, for experimental confrontations of CODYRUN (Lauret, 2001).

A weather station placed on the roof of the dwelling has recorded the meteorological datas (every 30 mn) : outdoor dry air temperature, sky temperature, relative humidity, global and diffuse horizontal radiation, speed and direction of the wind.

In each room, dry air and resultant temperature (in different locations), relative humidity, some surface temperatures and global radiation for the west side were measured.

The dwelling was uninhabited. Various measurement sequences were organized : windows closed, then partially open, during the extreme climatic period of the tropical summer.

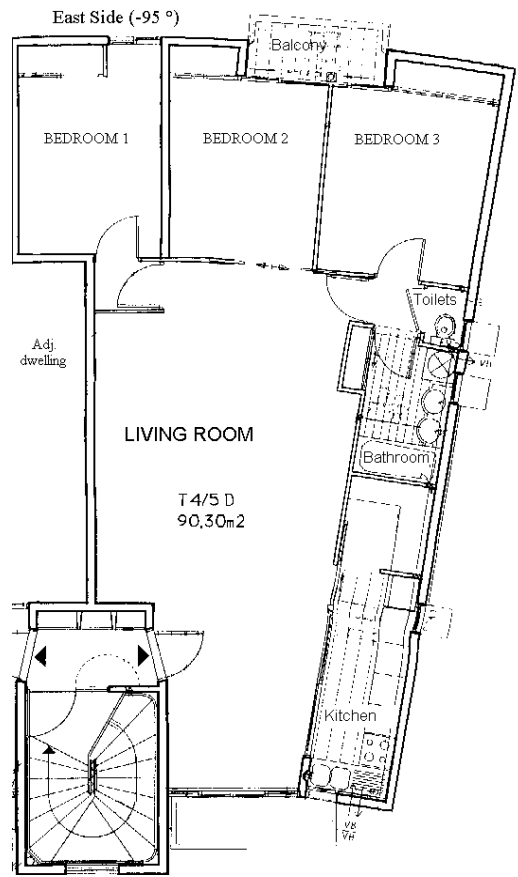

Fig. 4 : *Instrumented dwelling*

We consider two sequences:

**Closed building**

All the doors and windows of the dwelling were closed and sealed, during 7 days.

The building being closed, airflow model is of no use. The thermal model is time invariant.

The model of the dwelling is composed of five thermal zones. (the 3 bedrooms, the living room, the kitchen + bathroom + toilets).

Figure 5 shows the evolution of the dry temperature in the living room. The simulation is carried out by using the initial model. The figure shows the modelling error : the maximum difference between measures and the simulation results is about +/- 1.2 °C, and the standard deviation is 0.55 °C.

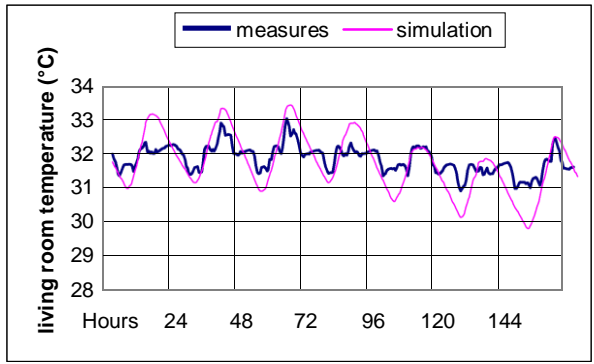

Fig. 5 : *comparison measures / initial model*

On figure 6 we compare a simulation results during 3 days, carried out by using first the initial model and then the reduced model.

The initial model consists of 5 sub-systems (i.e. 5 zones), orders are 33, 36, 37, 68 and 28.

The orders of the reduced models are automatically selected with a tolerance $\varepsilon = 0.2$. Orders become

Computing time is divided by 3.

Reduction errors (ie. the maximum difference between initial model result and reduced model result) are not significant, less than 0.2 °C.

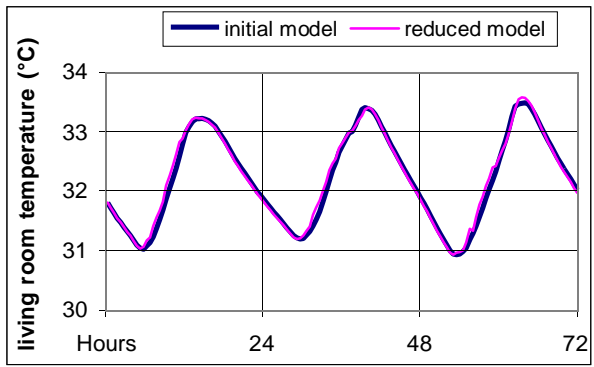

Fig. 6 : *comparison initial model / reduced model*

**Open building.**

The sliding door in the living room and window of bedroom 2 were open outwards, the sliding door between bedroom 2 and living room was open. Thus, air could circulate through the dwelling

In this case the thermal model is a time varying model. We used the separate model reduction method that we previously described.

On Figure 7 we compare dry temperature measured in the living room, and simulations carried out by using different methods :

- using the initial model.

It consists of 5 systems, orders are 38, 45, 48, 69 and 27.

- using the separate model reduction.

The order of the reduced models are automatically selected with a tolerance e = 0.4. Orders of the reduced systems become respectively 7, 6, 6, 11 and 5.

Because of the iterative procedure, the reduction of the computing time is less important than noticed previously. Computing time is divided by 2.

- using the conditional reduction.

Actually, this method is very slow for this case. Indeed, because of the natural airflow, the reduced model is very often computed.

The 3 simulated curves are very close.

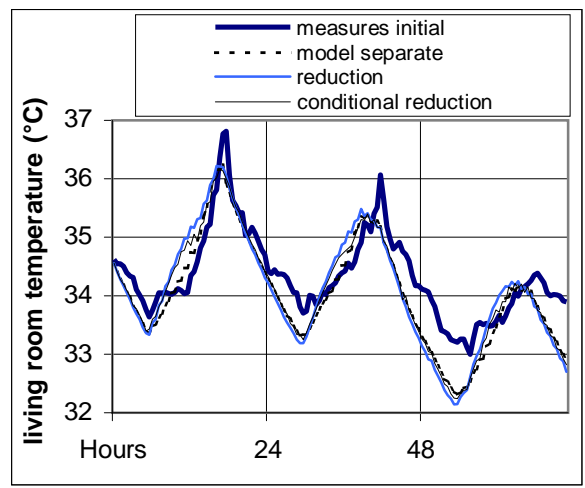

*simulations.*

## CONCLUSION

We implemented balanced model reduction within a software dedicated to thermal behavior and airflow simulation.

Hence, we are able to reduce half the global computing time required for a simulation**.**

Then, we compared experimental measures and simulation results using the full order model or the reduced model. The comparison shows very small reduction errors relatively to modelling errors.

Therefore, we have shown the effectiveness of balanced reduction tools for time varying systems.